\documentclass[10pt,conference]{IEEEtran}
\usepackage{cite}
\usepackage{amsmath,amssymb,amsfonts}
\usepackage{graphicx}
\usepackage{textcomp}
\usepackage[dvipsnames]{xcolor}
\usepackage{colortbl}
\definecolor{whiteOrange}{RGB}{255,200,100}
\usepackage[hyphens]{url}
\usepackage{fancyhdr}
\usepackage{hyperref}

\usepackage{comment}
\usepackage{booktabs}
\usepackage{subfig}
\usepackage{amsmath}
\usepackage{array} 
\newcolumntype{P}[1]{>{\centering\arraybackslash}p{#1}}
\newcolumntype{M}[1]{>{\centering\arraybackslash}m{#1}}
\usepackage{caption}
\usepackage{wrapfig}
\usepackage{algpseudocode}
\algrenewcommand{\algorithmiccomment}[1]{\hfill\textbf{//}\,#1}
\usepackage{algorithm}
\algrenewcommand\alglinenumber[1]{\small #1:}
\usepackage{circledsteps}
\pgfkeys{/csteps/inner color=white}
\pgfkeys{/csteps/fill color=black}
\usepackage{pifont}
\usepackage{multirow}
\usepackage{mdframed}

\usepackage{soul}

\usepackage{pdfcomment}

\newcommand{\hpcayear}{2024}

\edef\oldtt{\ttdefault}
\usepackage[scaled]{beramono}
\usepackage[T1]{fontenc}
\renewcommand{\ttdefault}{\oldtt}

\usepackage{mathtools}
\usepackage{booktabs} 
\usepackage[flushleft]{threeparttable} 
\usepackage{hyperref} 

\usepackage{setspace}
\usepackage{tikz}
\usepackage{pifont}
\usepackage{makecell}
\usepackage{pifont}

\title{ 
SAIL: SRAM-Accelerated LLM Inference System with Lookup-Table-based GEMV
}

\def\hpcacameraready{} 
\newcommand\hpcaauthors{Jingyao Zhang$^*$,  Jaewoo Park$\dagger$, Jongeun Lee$\dagger$, and Elaheh Sadredini$^*$}
\newcommand\hpcaaffiliation{University of California, Riverside$^*$\\
Ulsan National Institute of Science and Technology$\dagger$}
\newcommand\hpcaemail{}

\author{
  \ifdefined\hpcacameraready
    \IEEEauthorblockN{\hpcaauthors{}}
      \IEEEauthorblockA{
        \hpcaaffiliation{} \\
        \hpcaemail{}
      }
  \else
  \fi 
}

\fancypagestyle{camerareadyfirstpage}{%
  \fancyhead{}
  
  \fancyhead[C]{
    \ifdefined\aeopen
    \parbox[][12mm][t]{13.5cm}{\hpcayear{} IEEE International Symposium on High-Performance Computer Architecture (HPCA)}    
    \else
      \ifdefined\aereviewed
      \parbox[][12mm][t]{13.5cm}{\hpcayear{} IEEE International Symposium on High-Performance Computer Architecture (HPCA)}
      \else
      \ifdefined\aereproduced
      \parbox[][12mm][t]{13.5cm}{\hpcayear{} IEEE International Symposium on High-Performance Computer Architecture (HPCA)}
      \else
      \parbox[][0mm][t]{13.5cm}{}
    \fi 
    \fi 
    \fi 
    \ifdefined\aeopen 
      \includegraphics[width=12mm,height=12mm]{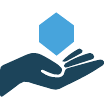}
    \fi 
    \ifdefined\aereviewed
      \includegraphics[width=12mm,height=12mm]{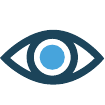}
    \fi 
    \ifdefined\aereproduced
      \includegraphics[width=12mm,height=12mm]{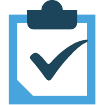}
    \fi
  }
  \fancyfoot[C]{}
}
\begin{document}
\maketitle

\ifdefined\hpcacameraready 
  \thispagestyle{camerareadyfirstpage}
  \pagestyle{empty}
\else
  \thispagestyle{plain}
  \pagestyle{plain}
\fi

\newcommand{\hpcaheight}{0mm}
\ifdefined\eaopen
\renewcommand{\hpcaheight}{12mm}
\fi


\begin{abstract}
Large Language Model (LLM) inference requires substantial computational resources, yet CPU-based inference remains essential for democratizing AI due to the widespread availability of CPUs compared to specialized accelerators. However, efficient LLM inference on CPUs faces two fundamental challenges: (1) existing CPU architectures struggle with low-precision arithmetic required by quantized models, where optimal bit precision varies across models and layers; and (2) the memory-bound nature of the token generation phase creates severe performance bottlenecks.
To address these challenges, we propose SAIL (SRAM-Accelerated Inference of LLMs), a CPU-based inference solution that efficiently supports arbitrary bit precisions with minimal overhead. SAIL integrates three key innovations: First, we introduce Batched LUT-based General Matrix-Vector Multiplication (LUT-GEMV) with SRAM-based processing-in-memory, enabling high data reuse through lookup tables and reducing memory movement. Second, our Pattern-Aware LUT optimization identifies and exploits redundancy in input activation patterns, reducing computation cycles by 13.8\%. Third, we develop an in-memory type conversion algorithm that leverages PIM's parallelism for efficient de-/quantization operations, alleviating pressure on CPU's vector units.
Our architecture requires only 2\% hardware overhead and a single new instruction, while maintaining dual functionality as both compute and storage units. Experimental evaluations using a modified gem5 simulator demonstrate that SAIL achieves up to 10.7× speedup and 19.9× higher tokens per dollar compared to ARM Neoverse-N1 CPU baselines, and up to 7.04× better cost efficiency than NVIDIA V100 GPUs, establishing a practical path for efficient CPU-based LLM inference.
\end{abstract}

\section{Introduction}\label{intro}

Large Language Model (LLM) inference presents fundamental computational challenges that extend beyond traditional AI workloads \cite{Seger2023-xy,Ahmed2020-nk}. The token generation phase of LLM inference is inherently memory-bound, requiring continuous access to model weights that far exceed on-chip memory capacity \cite{Wang2024-lr,Menshawy2024-ao}. While GPUs dominate AI training, their limited availability, high costs, and regulatory constraints create barriers to widespread LLM deployment \cite{Zhao2024-bd,Jeltes2020-mf}.

CPUs offer a compelling alternative for democratizing AI inference due to their widespread availability and lower barriers to deployment \cite{Ditto2024-df,Zeng2024-ql}. Recent CPU architectures have shown promising results for AI workloads, demonstrating that efficient inference is possible without specialized accelerators \cite{Zhang2024-ek,Ashfaq2022-im}. This makes CPU-based inference particularly attractive for organizations and regions with limited access to specialized hardware \cite{Kamahori2024-re,Gholami2024-rh}.

However, efficient LLM inference on CPUs faces two critical challenges. First, model quantization—essential for reducing memory bandwidth requirements—demands flexible low-precision arithmetic that varies across models and even layers within the same model \cite{noauthor_undated-fn,Dumitru2024-mn}. Current CPU architectures struggle with sub-8-bit operations, experiencing significant inefficiencies when processing 2-4 bit quantized weights \cite{Park2024-zw,Verma2023-nb}. Second, the memory-bound nature of token generation creates severe bottlenecks, with data movement dominating execution time \cite{Gong2024-mv,Hu2024-ig}.

Processing-in-Memory (PIM) has emerged as a promising solution to address the performance and energy efficiency challenges in AI inference tasks. Existing research has shown that PIM architectures utilizing bit-serial-based computing can outperform GPUs in certain scenarios \cite{Fujiki2019-fi}.
However, when these approaches are applied to quantized Large Language Models (LLMs), significant limitations become apparent.

\begin{figure}[tp]
\centerline{\includegraphics[width=2.6in]{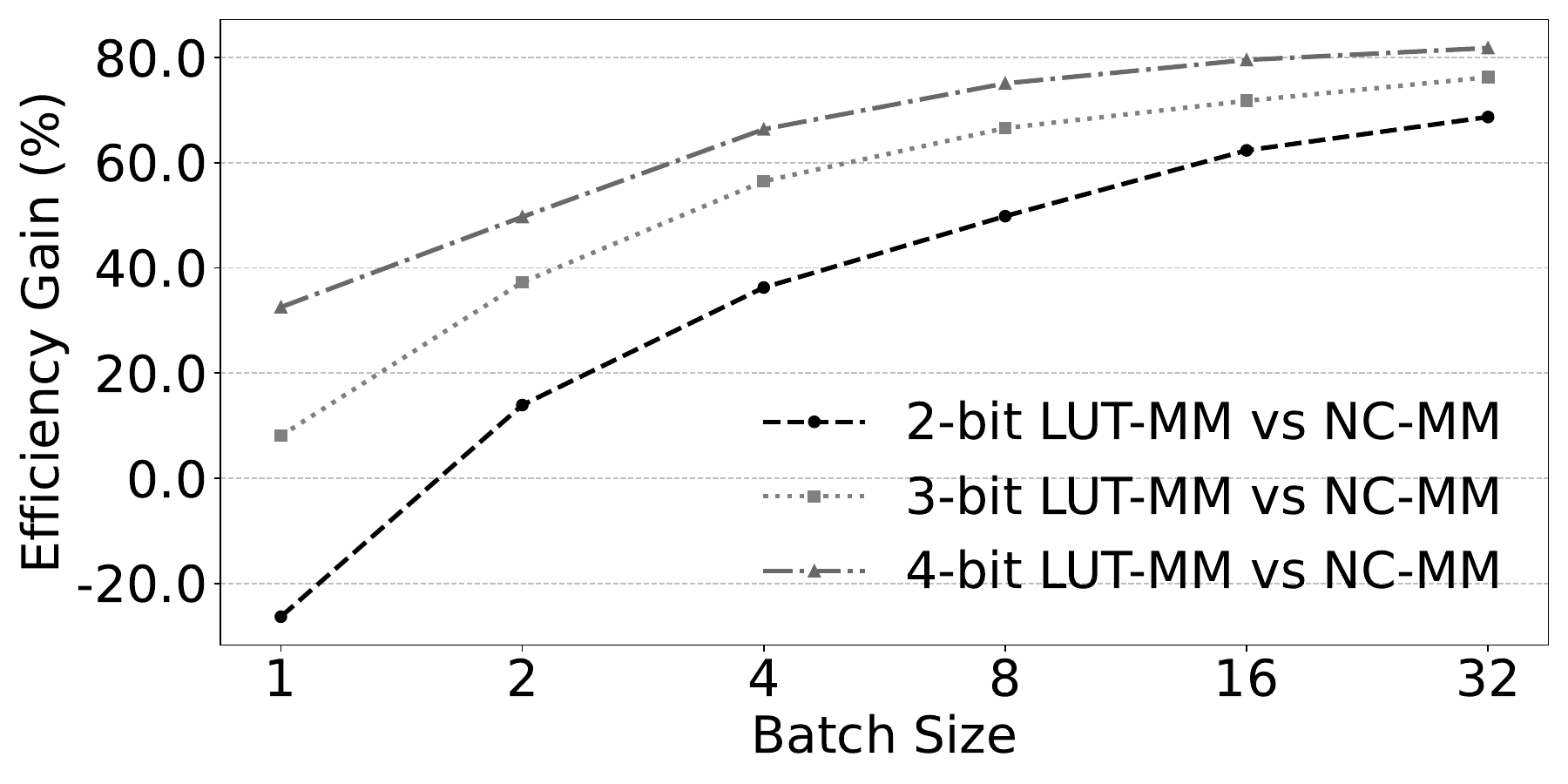}}
\caption{
Efficiency gain comparison between LUT-based and bit-serial computing \cite{eckertNeuralCacheBitSerial2018a} for 2-bit, 3-bit, and 4-bit quantization across various batch sizes. Dashed lines represent different bit-width quantizations.
}
\label{fig:effi}
\end{figure}

Our investigation reveals two critical findings in this context:
First, \textbf{LUT-based computing demonstrates superior performance over bit-serial computing in low-precision model inference}. LUT-based methods capitalize on the efficiency of lookup tables to handle computations at low bit precisions, as shown in Fig. \ref{fig:effi}.
Second, \textbf{current PIM designs often introduce substantial overheads or increased latency during system integration}. Up to 90\% of the processing time may be spent waiting for data to move to specific cache locations \cite{Al-Hawaj2023-rr}, which severely diminishes the advantages gained from in-memory computation.
These findings highlight the necessity for a new PIM architecture that can mitigate these limitations. By focusing on LUT-based computing and optimizing data movement within the system, we can develop a more efficient PIM solution for quantized LLM inference.

Our proposed solution, SAIL, features the following architecture- and system-level innovations. (1) Observing that the internal bandwidth among Last Level Cache (LLC) slices is often underutilized, we propose to add a compute-capable SRAM (called \emph{C-SRAM}) array adjacent to each cache slice, enabling direct data transfer within the cache and very high data bandwidth to C-SRAM. The C-SRAM array performs in-place computing in the manner of LUT-GEMV, which has low cost and arbitrary precision support.
(2) At the same time, our architecture supports flexible data placement in caches and external memories, crucial for supporting very large workload and pipelining between compute and data transfer. This motivates us to propose tensor-level scheduling and pipelining for LLM inference.
(3) We also propose a novel in-memory type conversion algorithm that tightly couples LUT-GEMV with existing compute and memory systems.  SAIL efficiently handles floating-point computation by orchestrating high parallelism in PIM and leveraging the CPU vector engine for efficient floating-point operations.  (4) Finally, to natively support quantized LLM (QLLM) inference, we introduce a new instruction with fields for GEMV scale factors and corresponding quantization levels. These architectural enhancements allow for minimal system integration overhead while supporting flexible precision in inference operations.

Our experimental evaluations, conducted using a modified gem5 simulator, demonstrate that SAIL outperforms both CPU and GPU baselines in terms of speed and cost efficiency for AI inference. Leveraging its architectural innovations, SAIL achieves notable improvements in throughput and tokens per dollar, highlighting its potential to advance CPU-based AI inference and contribute to the broader goal of AI democratization (see Section~\ref{tpd-def} for cost estimation methodology). We focus on a multi-user scenario where multiple users request LLM inference services deployed on servers, a common use case for which our proposed architecture is well-suited.


Our key contributions are:
\begin{itemize}
\item \textbf{Near-cache PIM architecture:} We integrate LUT-based GEMV with cache-adjacent C-SRAM arrays, enabling efficient arbitrary-precision arithmetic while maintaining dual compute/storage functionality.
\item \textbf{System-level optimizations:} We develop tensor-level scheduling and pipelining techniques that overlap computation with data movement, along with a Pattern-Aware LUT optimization that reduces computation cycles by 13.8\%.
\item \textbf{Hardware-software co-design:} We propose an in-memory type conversion algorithm and a single new instruction that enables complete QLLM inference support with only 2\% hardware overhead.
\item \textbf{Comprehensive evaluation:} We demonstrate 10.7× speedup and 19.9× better cost efficiency compared to CPU baselines, with competitive performance against GPUs at much lower cost.
\end{itemize}

\section{Background} \label{bg}

\subsection{Inference and Quantization of LLMs}
\label{sec:llm_inference_quant}

Recent progress in Large Language Models (LLMs) such as GPT~\cite{yenduri2023generativepretrainedtransformercomprehensive} has popularized decoder-only, auto-regressive architectures in generative tasks~\cite{Minaee2024-lj}. LLM inference typically involves two stages: a \emph{prompt processing} (prefill) stage that uses matrix--matrix multiplication in parallel (compute-bound), and a \emph{token generation} stage that iteratively generates tokens with matrix--vector multiplication (memory-bound). While prompt processing occurs only once, the token generation phase repeats until the full output sequence is produced, often dominating latency in LLM inference. To reduce recomputation, intermediate Key and Value activations (KV cache) are stored~\cite{Duanmu2024-be,Gerganov_undated-gt}, but their size scales linearly with the sequence length, batch size, and number of layers. For models like Llama-2-7B~\cite{Touvron2023-qn} at FP16 precision and context length 4096, the KV cache can reach or exceed the size of the model weights, making it challenging to leverage large batch sizes on current GPU VRAM~\cite{Kwon2023-bh}.

Quantization of LLMs is a common strategy to reduce the model size and memory footprint~\cite{Lin2023-nk,fasterxf-jl,Dettmers2022-tz,liu2023llm,Xiao2022-qg}, but it remains difficult on existing hardware due to low-precision arithmetic and mixed-precision weight/activation schemes. In contrast to CNN quantization approaches~\cite{Krishnamoorthi2018-dp,esser2019learned,choi2018pact}, sub-8-bit LLM quantization often requires intricate methods like group-wise or mixed-precision quantization~\cite{Xiao2022-qg}. A small fraction of outlier weights may even remain at higher precision to preserve accuracy in larger models. GPUs and CPUs, optimized for high-throughput floating-point or integer operations, experience significant \emph{data redundancy} when performing sub-8-bit computations; for instance, a 128-bit vector engine may only use 72 effective bits in a 4-bit $\times$ 8-bit scenario, leading to underutilization of compute lanes~\cite{Gerganov_undated-gt}. Moreover, activation quantization further complicates hardware support for LLMs, as most ASIC-based DNN accelerators lack built-in flexibility for these specialized quantization formats. 
Look-Up Table (LUT) computing~\cite{Lee2019-ac} offers a potential remedy by mapping low-precision multiplications to simple table-lookup and addition operations. This approach can eliminate data redundancy and handle complex quantization schemes with minimal hardware changes. Such near-data or in-SRAM LUT operations are especially appealing for the memory-bound token generation stage in LLMs, since they can reduce data movement and achieve high parallelism without incurring the overheads common in CPU or GPU vector units.

\subsection{Processing-in-Memory}\label{sec:insram}
Processing-in-Memory (PIM) technology offers a promising solution, providing high computational parallelism and effectively addressing the memory wall problem \cite{Asifuzzaman2023-ix}. However, existing PIM approaches can be categorized into three types, each with its own limitations:
\textit{1) Off-chip PIM} (e.g., DRAM-based PIM, SSD-based PIM) \cite{Gomez-Luna2021-mu,Gao2021-sm}: While offering the highest parallelism, these solutions may struggle when the PIM hardware and the CPU must cooperate frequently for computations that cannot be done easily by PIM, e.g., handling of \textbf{floating-point computations required for de-/quantization operations} in LLMs, which can account for approximately 50\% of quantized LLM inference workloads \cite{Kim2023-st}.
\textit{2) On-chip in-cache computing} \cite{Al-Hawaj2023-rr,Aga2017-kx,Eckert2018-cl}: This approach enables efficient floating-point computations by involving the CPU through on-chip data movement. However, it either requires \textbf{extensive OS and memory allocator modifications} \cite{Aga2017-kx,Eckert2018-cl} or introduces \textbf{additional latency} \cite{Al-Hawaj2023-rr} due to lengthy cache-internal data movements (up to 90\% of time spent waiting for data to move to specific cache locations).
\textit{3) On-chip near-cache computing} \cite{Wang2022-nq,Wang2023-az,Reis2022-uv}: While avoiding substantial software overhead and data movement costs, this approach is constrained by \textbf{cache external bandwidth}, resulting in suboptimal performance.

To efficiently support quantized LLM (QLLM) inference, these questions must be answered: how to scale compute units, and how to provide high enough data bandwidth for the compute units. Also, certain operations of QLLM inference cannot be done easily by PIM.
Full integer quantization of QLLM is infeasible, mandating support for both floating-point and low-bit integer operations \cite{Lin2023-nk,Xiao2022-qg}.
Further, LLMs deal with large amount of data; weights may not fit in on-chip caches, and activations including KV cache \cite{hooper2024kvquant10millioncontext} are often even larger than weights, making pipelining between compute and data transfer inevitable. Traditional in-memory computing methods are too rigid in terms of data placement to support pipelined execution. Near-memory computing is more flexible but suffers from lower data bandwidth, and its compute units (e.g., ALUs/MACs) are rather costly and inefficient on low-precision arithmetic.

\begin{figure}[htp]
\centerline{\includegraphics[width=2.8in]{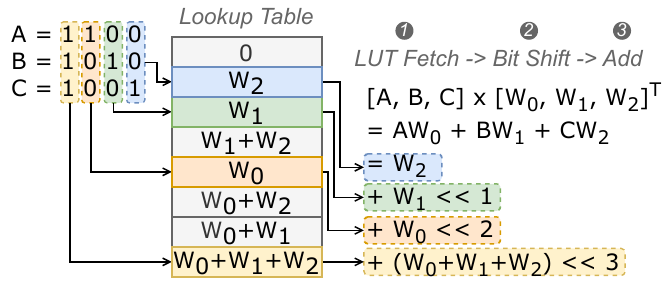}}
\caption{
LUT-based vector multiplication for a 4-bit input vector $[A, B, C]$ and weights $[W_0, W_1, W_2]$ using bit-serial computation with NBW = 3.
}
\label{fig:lut}
\vspace{-1em}
\end{figure}

\subsection{Low-precision Matrix Multiplication Using Lookup Tables}

Matrix multiplication is a cornerstone operation in numerous computing tasks, particularly in the domains of machine learning and signal processing. The efficiency
of this operation is paramount, especially when dealing with vast amounts of low-precision data from quantized models. Common platforms (e.g., CPU and GPU) do not efficiently deal with low-precision computation other than their natively supported data formats.
A promising solution to efficiently handle various data precisions is
Lookup Table (LUT)-based matrix multiplication.
The LUT approach \cite{Lee2019-ac}
offers a cost-effective construction process, and once created, the LUT can be utilized repeatedly, reducing computational overhead.
Fig.~\ref{fig:lut} illustrates the process for vector multiplication, which serves as a foundational concept extendable to matrix multiplication.
Consider a scenario where we have three $n$-bit weights denoted as $W_0$, $W_1$, $W_2$, accompanied by three 4-bit activation inputs labeled $A$, $B$, $C$. The objective is to compute the expression $AW_0+BW_1+CW_2$. By employing a LUT of weights and adhering to a bit-serial processing of input values, the task is simplified to just four steps of shift-and-adds. As shown in Fig.~\ref{fig:lut}, the LUT is constructed based on 3 weights, which we define as the \textbf{Number of Basis Weights (NBW) = 3}. This term specifies the number of basis weights used for LUT construction, resulting in a total of $2^3 = 8$ entries.

The construction of the LUT is the first phase, predicated on the given weights. In our case, three weights necessitate a LUT with $2^3$ entries. Generally, for $k$ weights, the LUT would require $2^k$ entries. This pre-computed table, which can be constructed using the bitline-style in-SRAM computing (see the previous section), allows for rapid retrieval of weight combinations based on the activation inputs' bit patterns.
Subsequently, we inspect the activation inputs' bits, progressing from the least significant bit (LSB) to the most significant bit (MSB). Each bit pattern corresponds to a LUT entry, which is then used to fetch the relevant weight-sum. For instance, the bit pattern \texttt{001} retrieves $W_2$ from the LUT, as shown in Fig.~\ref{fig:lut}. As we advance to higher-order bits, the corresponding weight is left-shifted to align with its bit position before being added to the ongoing sum. This ensures that each bit's contribution is accurately reflected in the final result.
The procedure continues iteratively: the second least significant bits prompt a fetch of $W_1$, which is then left-shifted by one bit position. As we proceed to higher bit positions, we fetch $W_0$ for \texttt{100} and $W_0+W_1+W_2$ for \texttt{111}, each followed by a corresponding left shift. These values are then added to the partial sum to incrementally build the final result.


\subsection{Design for LUT-GEMV in PIM}

LUT-based GEMV integration into PIM architectures for quantized LLM inference faces critical optimization challenges. The Number of Basis Weights (NBW) parameter creates a fundamental trade-off: increasing NBW exponentially grows LUT size (2$^{NBW}$ entries), while decreasing NBW requires more computational rounds. Current implementations lack systematic analysis of this trade-off in PIM contexts.
Our preliminary investigations revealed significant input pattern reuse across computations—a characteristic current designs fail to exploit. This unexplored opportunity motivates our systematic analysis of optimal NBW configurations and pattern reuse techniques to overcome memory-compute tensions in LUT-GEMV implementations for quantized LLM inference. Prior LUT-based accelerators, such as Stella Nera \cite{Schonleber2025-in} and T-MAC \cite{Wei2025-wh}, either operate in software or as standalone units. In contrast, SAIL is the first to co-locate LUT-based compute with storage, which eliminates data movement, enables per-token dynamic LUTs, and allows the hardware to function as a normal cache when idle.

\section{Our Approach: \textit{SAIL}} \label{sec:arch}

To address the challenges associated with low-precision matrix multiplication, type conversion performance, and the memory-bound issues in the LLM decoding stage, this paper introduces the \textit{SAIL (SRAM-Accelerated
Inference of LLMs)} framework. Overall, this framework leverages near-memory computing to alleviate the memory wall problem, utilizes Lookup Tables to efficiently perform low-precision matrix multiplication, and employs in-memory type conversion to bridge the high parallelism of in-SRAM computing with the high-performance floating-point processing capabilities of the CPU vector engine. 
\emph{In contrast with prior SRAM-based solutions \cite{Eckert2018-cl, Aga2017-kx}, which may be called \emph{in-cache computing}, a distinguishing feature of SAIL (which is \emph{near-cache computing}) is its utilization of cache as a ping-pong buffer to establish a pipeline, rather than leveraging data locality.} The results demonstrate that this pipelining strategy enhances the acceleration of LLM inference.

\begin{figure}[htp]
\centerline{\includegraphics[width=3.0in]{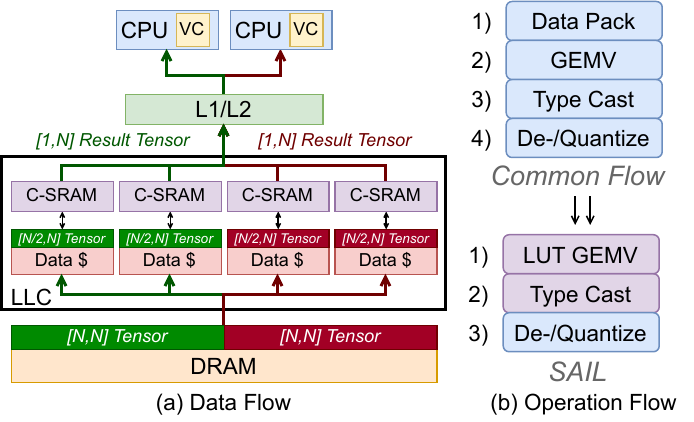}}
\caption{
(a) Data flow of SAIL. (b) Operation of common CPU-based inference vs. SAIL.
}
\label{fig:flow}
\end{figure}

In the proposed SAIL, as illustrated in Fig.~\ref{fig:flow}(a), the workflow initiates with tensors being retrieved from DRAM and stored in data cache slices. The C-SRAM modules then fetch the data segments to carry out LUT-based GEMV, subsequent to which the results are directed to the CPU for de-/quantization.
Contrasting the current CPU-based inference flow (as shown in Fig.~\ref{fig:flow}(b)), which necessitates the transfer of entire tensors through LLC external bandwidth---typically two $[N,N]$ tensors---SAIL reduces this requirement by only transferring result tensors, which are two $[1,N]$ tensors. 
Moreover, SAIL leverages the high parallelism afforded by in-SRAM computing, which not only expedites the GEMV but also enhances the efficiency of type casting.

\subsection{Tensor-level Scheduling and Pipeline Design}

We propose tensor-level scheduling to enhance the temporal locality of weights, thereby reducing the overhead associated with data movement. Notably, unlike previous work on in-SRAM computing \cite{Eckert2018-cl,Aga2017-kx}, we explicitly consider the costs of moving data from DRAM to cache and propose a novel strategy to minimize these expenses. Typically, inference serving systems operate on an iteration-based principle when serving multiple users \cite{Yu2022-ri,Kwon2023-bh}, meaning that each request from a user triggers the computation of the entire model (for example, a 32-layer Llama-2-7B model) followed by another user's computation. Due to cache capacity limitations, no current CPU cache can load the entire model at once. 
Therefore, we propose tensor-level scheduling, which involves loading the weights of one layer into the LLC cache at a time, and then processing this tensor's computations for different users, thus increasing the temporal locality of layer weight parameters. As a result, during a batched inference session serving multiple users, the model weights are loaded from DRAM to cache only once, optimizing performance and reducing data movement overhead.

\begin{figure}[htp]
\centerline{\includegraphics[width=3.4in]{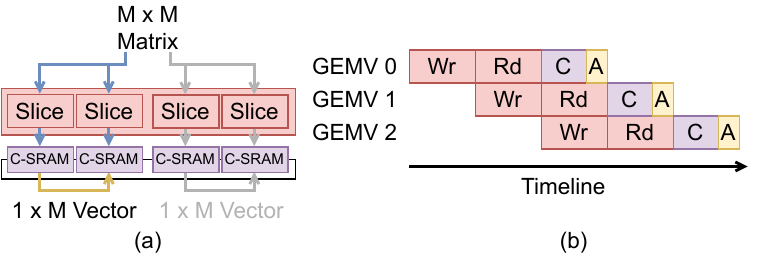}}
\caption{
(a) Detailed diagram of computation flow over the proposed architecture. With ping-pong cache, matrix with size of M$\times$M is written into one half of the cache and then read by the C-SRAM. The C-SRAM perform computation and aggregation of partial results to generate the final results of GEMV. (b) The pipeline diagram. The designed pipeline can be full without bubbles. The write, read and computation (including aggregation) can be fully overlapped.}
\label{fig:pipeline}
\end{figure}

During our experiments, computational time is much less than the time required for data movement. This discrepancy is largely because the decoding stage is memory-bound, meaning that even attempts to overlap computation with data movement cannot achieve complete overlap. However, our adoption of tensor-level scheduling reduces the frequency of weight data transfers and enhances temporal locality, diminishing the overhead associated with data movement. Furthermore, we can identify the most suitable batch size (the number of users served per unit of time) to balance the entire pipeline. Thus, we propose using the LLC as a ping-pong buffer to overlay computation and data movement. Specifically, we divide the LLC cache slices into two parts: while the first part is being written with weight tensors, the second part outputs tensors to the C-SRAM for processing and writes the computation results its data cache. In the subsequent phase, the roles of the two parts are reversed, thereby creating a ping-pong buffer effect that overlaps data movement and computation. After measuring the computational and data movement costs, we found that the pipeline performs optimally when the batch size is 8. However, this analysis only considers the pipeline's design and not the overhead of the LUT-based GEMV. In Section \ref{sec:dse}, we will discuss the performance and costs associated with batch size and LUT-based GEMV, to determine the optimal batch size.

\subsection{Compatibility with KV-cache}

\begin{figure}[htp]
\centerline{\includegraphics[width=3.0in]{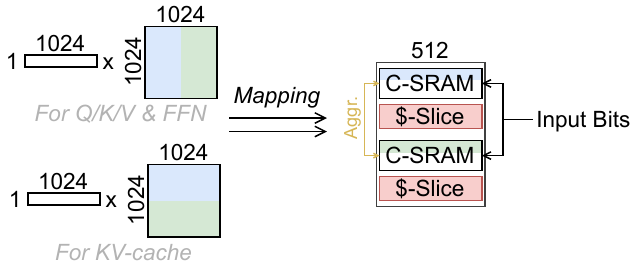}}
\caption{
Diagram of matrix mapping over compute SRAM. For common matrix multiplication (e.g., Q/K/V and feed-forward layer), weights at the same row are split into different C-SRAM arrays. For KV-cache related computation (e.g., one vector multiplied by a transposed matrix built on KV-cache entries), weights at the same column are split into different C-SRAM arrays.}
\label{fig:allocation}
\end{figure}

The KV-cache is a commonly used technique to accelerate LLM inference \cite{Strati2024-cg, Kang2024-zr, Hooper2024-oc, Dong2024-hp}. During the decoding stage, computations are carried out only for the embedding of a single token generated from the previous iteration, while previously computed KV entries stored in the KV-cache are fetched to complete all necessary computation. With a standard KV-cache, each new token produces a vector (e.g., dimension d=4096 in Llama-2) that is multiplied by the cached K or V matrix (shape Txd, where T is the context length). Thus, only one input vector changes per iteration; the cached matrix is static during that iteration.

Our proposed method supports both quantized and non-quantized KV-caches. After each LUT-based GEMV is completed, the output (e.g., a key entry) is sent to the CPU’s vector engine for dequantization. At this point, if a quantized KV-cache is used, the CPU's vector engine performs an additional light-weight re-quantization step according to the KV-cache's quantization requirements before storing it in memory. If a non-quantized KV-cache is used, the output is stored directly in memory. The CPU only processes a single vector per token, making the load negligible.

Profiling shows that KV-related dynamic matrix multiplication, such as the Q x K\_cache\textsuperscript{T} operation, accounts for approximately 5\% of the total end-to-end latency, meaning LUT-GEMV remains the dominant computation. As illustrated in Fig. \ref{fig:allocation}, we map the transposed KV matrix column-wise across C-SRAM arrays. This allows the vector-matrix product to stream efficiently through the arrays without rebuilding large LUTs. In summary, the KV-cache path uses the same hardware as weight GEMV with minimal overhead. We will add this performance breakdown to the camera-ready paper.

\subsection{Design Space Exploration}\label{sec:dse}

\begin{figure*}[tp]
\centerline{\includegraphics[width=6.8in]{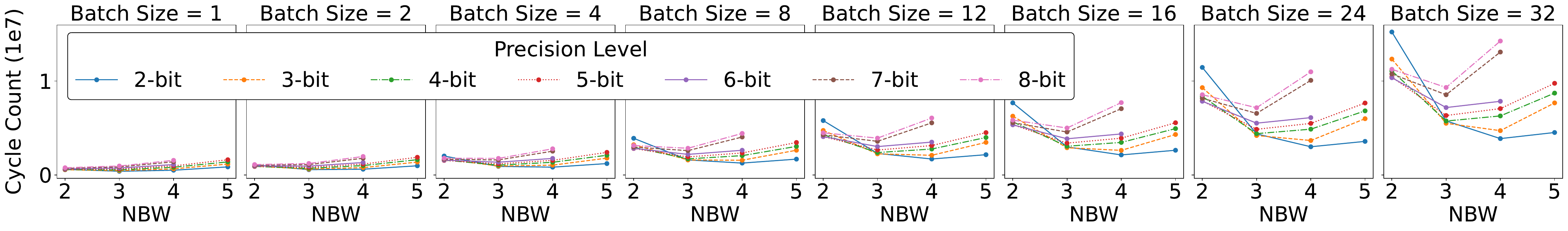}}
\caption{
Cycle count trends for different precision levels and NBW (Number of Basis Weights) values across various batch sizes. The optimal NBW varies depending on the batch size and precision level.}
\label{fig:NBW_cycle_count_trends}
\vspace{-1em}
\end{figure*}

Figure~\ref{fig:NBW_cycle_count_trends} illustrates how the cycle count changes under varying \textbf{batch size}, \textbf{precision level}, and \textbf{number of basis weights (NBW)}. At smaller batch size (e.g., 1 or 2), the cycle count remains high regardless of NBW, because the overhead of LUT creation is not effectively amortized. As the batch size grows, the cycle count drops substantially but plateaus beyond about 7, indicating limited returns for further increases.
A larger NBW exponentially increases the LUT size and can reduce the cycle count slightly. However, it also diminishes the \textit{Arithmetic Intensity} (the ratio of computation to memory traffic) when the batch size is small. In such cases, fetching a large LUT quickly saturates \textbf{memory bandwidth}, providing fewer operations per byte fetched and producing little net speedup. At higher batch sizes, moderate NBW values (e.g., 2 or 3) offer a better trade-off between LUT overhead and execution latency. The precision level interacts with these factors as well: lower-precision quantization (Q2 or Q3) can magnify LUT growth and memory bandwidth concerns, whereas higher-precision models (Q6 or Q8) incur greater arithmetic cost but may be more bandwidth-friendly.
In addition to these runtime considerations, a final factor is whether the LUT is \textbf{constructed offline/online}. Offline LUT construction precomputes all tables before inference, removing setup costs at runtime but inflating the model size (by up to 3.75$\times$ at Q4 with NBW=4). In memory-rich environments, this trade-off may be acceptable for reduced latency. In this work, however, we opt for online LUT construction to avoid prohibitive model-size expansion and to maintain precision flexibility. This approach creates LUTs on-the-fly and reuses them within the batch, providing a more balanced solution. The overhead of online LUT creation ranges from 3\% (batch=8, NBW=2, 2-bit) to 12\% (batch=32, NBW=4, 4-bit) of end-to-end inference cycles, an acceptable cost for the memory savings and flexibility gained.

SAIL fully supports arbitrary weight precision. The maximum precision in one C-SRAM column is determined by $\text{bit\_width}_{\max} = \lfloor R / 2^{\text{NBW}} \rfloor$, where $R$ is the row count (256 in our prototype). With NBW=2, we can theoretically support up to 64-bit weights. In practice, LLM inference uses $\le$8-bit precision, which is fully covered in our evaluation. Reducing bit-width generally lowers latency, provided NBW and batch size are co-optimized. For instance, at a batch size of 24, using NBW=4 with 2-bit precision requires 3.00M cycles, while 4-bit precision takes 4.87M cycles. However, simply lowering precision is insufficient; with NBW=2, 2-bit inference takes 11.45M cycles due to LUT rebuild overhead. Thus, SAIL jointly optimizes the {NBW, bit-width, batch size} design space to maximize speedup. We will add a figure illustrating this trade-off in the final manuscript.

\subsection{Pattern-Aware LUT Optimization}

Our analysis of LLM inference workloads revealed that approximately 17\% of input activation patterns repeat within computation batches. To exploit this redundancy, we implemented a Pattern-Aware LUT Access mechanism. Each Data Feeding Module (DFM) contains a 32-entry fully-associative Pattern Reuse Table (PRT). The PRT stores a 32-bit hash of the NBW-bit input pattern along with the previous LUT result. On a PRT hit, the DFM bypasses the C-SRAM access and reuses the stored result. This optimization reduces computation cycles by 13.8\%.

The hardware cost is minimal. Synthesized in FreePDK-45nm, a single PRT, including its 16-bit adder tree for merging partial sums, occupies approximately 0.0012~mm\textsuperscript{2} and consumes 0.25~mW. For a system with eight DFMs, the total hardware overhead is less than 0.01~mm\textsuperscript{2} (under 0.08\% of a 32MB LLC) and under 2~mW.

\subsection{In-Memory Parallel Type Conversion}
\label{subsec:in_memory_conversion}

We propose a new in-memory parallel algorithm (Algorithm \ref{alg:conversion}) to convert signed integers to IEEE-754 single-precision floating-point numbers, a process that is both crucial and compute-intensive in many applications.
Given an $n$-bit signed integer $A$ ($n \le$ 25), the algorithm converts the integer into a 32-bit floating-point number $R$, conforming to the IEEE-754 standard format.\footnote{The other direction (i.e., FP to INT) is straightforward. Our algorithm does not support exceptional cases such as NaN and subnormal numbers.} The conversion process is meticulously crafted to be fulfilled using in-memory computing only, utilizing logical operations to minimize data movement and accelerate the conversion process.
The algorithm begins by iterating over the bits of $A$, during which a 1-bit variable $D$ is used to find and record the position of the first `1' in $A$, constructing an $(n-1)$-bit number $C$, where all the bits following the first `1' are set to `1'. Then the exponent can be found by simply counting the number of 1s in $C$. The mantissa is obtained by multiplying $A$ with $2^k$, where $k$ is the number of leading zeros in $A$ (Lines \ref{bitreverse}-\ref{conv_mult}).
The algorithm requires $O(n^2/2 + 13(n-1))$ logical operations, which can be performed in $(3n^2/2+39(n-1))$ cycles in in-SRAM computing.
Our novel algorithm not only streamlines the type conversion process but also aligns with the growing trend of in-memory computing, where computational tasks are increasingly integrated with memory subsystems.

\begin{algorithm}[htp]
\small

\caption{In-memory parallel type conversion to float}
\label{alg:conversion}
\begin{algorithmic}[1]
\Require $n$-bit signed integer $A=(a_{n-1},...,a_0)$, where $n \leq 25$
\Ensure 32-bit IEEE-754 single-precision floating-point number $R$

\State $C:=(c_{n-2},...,c_0)=0$; $D:=0$; $R:=(r_{31},...,r_0)=0$; $Sum:=(s_4,s_3,s_2,s_1,s_0)=0$

\For{$i=n-2$ to $0$}
    \State $D := D \,|\, a_i$; $c_i := c_i \,|\, D$ \algorithmiccomment{Find leading 1 and record position}
\EndFor

\For{$i=0$ to $n-2$}
    \State $\texttt{Carry}:=c_i$
    \For{$j=0$ to $4$}
        \State $\texttt{\textit{c1}} := s_j \,\&\, \texttt{Carry}$; $s_j := s_j \oplus \texttt{Carry}$; $\texttt{Carry} := \texttt{\textit{c1}}$
    \EndFor
\EndFor

\State $\texttt{Sum} := \texttt{Sum} + 126$ \algorithmiccomment{Compute biased exponent}
\State $r_{31} := r_{31} \,|\, a_{n-1}$ \algorithmiccomment{Set sign bit}

\For{$i=23$ to $27$}
    \State $r_i := r_{i} \,|\, s_{i-23}$ \algorithmiccomment{Set biased exponent}
\EndFor

\State $C := \mathrm{BitReverse}(C + 1) << 1$  \label{bitreverse} \algorithmiccomment{Bit-serial operations}
\State $A := A * C $   \label{conv_mult} \algorithmiccomment{Align mantissa bits}

\For{$i=0$ to $n-3$}  \algorithmiccomment{Set mantissa, remove hidden 1}
    \State $r_{22-(n-3)+i} := r_{22-(n-3)+i} \,|\, a_{i}$
\EndFor

\end{algorithmic}
\end{algorithm}

\section{Implementation}

In this section, we introduce our implementation of SAIL, which supports the proposed tensor-level scheduling strategy designed to merge batched inference with LUT-based GEMV.

\subsection{ISA Design}

\begin{figure}[htp]
\centerline{\includegraphics[width=3.0in]{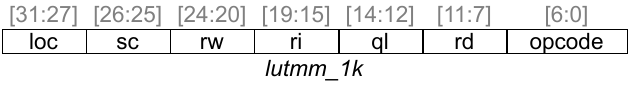}}
\caption{
Extension of RISC-V instruction set architecture.
}
\vspace{-1em}
\label{fig:isa}
\end{figure}

To support our proposed LUT-based in-SRAM computing model, we extend the RISC-V ISA as shown in Fig.~\ref{fig:isa}. The new instruction is designed to perform a tiled GEMV computation of size [1,1024]$\times$[1024,1024] using LUT-based in-SRAM computing. We choose to support a size of 1024 because most current models have hidden sizes and feed-forward dimensions that are multiples of 1024 (for example, the hidden size for OPT-350M \cite{Zhang2022-wg} is 1024, $ffn\_dim$ is 4096; hidden size for Llama-2-70B \cite{Touvron2023-qn} is 8192, $ffn\_dim$ is 28672). This enables the computation of GEMVs such as [1,1024]$\times$[1024,4096] by piecing together the computations of [1,1024]$\times$[1024,1024]. For sizes that are not exact multiples, one approach is to use padding, and another is to introduce new instructions. In this paper, we focus on GEMV computations with a dimension of 1024.

As illustrated in Fig.~\ref{fig:isa}, the instruction includes several fields: the location ($loc$) of the tile GEMV within the full GEMV computation, the scale factor ($sc$) relative to the size of 1024, the register index storing the starting address of the weights ($rw$), the register index for the starting address of the input ($ri$), the quantization level ($ql$) for the computation, the target register index storing the starting address of the result vector ($rd$), and the opcode. Using $loc$ and $sc$, we can determine how to fetch the required [1024,1024] weight tile from the complete weight matrix. For instance, if $sc$ is 3, it implies the entire weight matrix's width is 1024$\times$2$^3$=8192, and if $loc$ is 5, the required [1024,1024] weight tile would be from column index 5$\times$1024=5120 to 6$\times$1024=6144. Accordingly, the corresponding 1024-sized weight matrix can be fetched using the starting address stored in $rw$. The starting address stored in $ri$ will then be broadcasted by the data feeding module to all C-SRAM arrays, and the address of the final computation result will be stored in $rd$. $ql$ indicates the quantization levels. Our LUT-based method can support computations of various precisions, thus we currently support all common quantization levels (2/3/4/5/6/8-bit).

\subsection{Overall Architecture}

\begin{figure*}[tp]
\centerline{\includegraphics[width=6.6in]{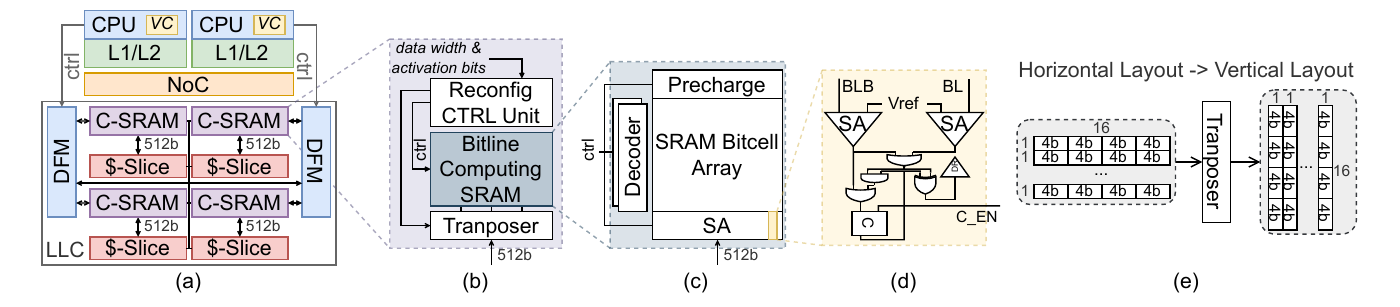}}
\caption{Overview of our reference implementation. (a) A scalable multicore with shared LLC and individual private caches, extended with C-SRAMs and Data Feeding Modules, (b) C-SRAM, (c) BC-SRAM, (d) Modified SAs, and (e) Transposer.
}
\label{fig:overall}
\vspace{-1em}
\end{figure*}

Our reference implementation, illustrated in Fig.~\ref{fig:overall}, introduces a scalable compute approach by integrating C-SRAMs with the fabric of the last-level cache (LLC). While the diagram shows a dual-core configuration, the architecture is designed to support an unlimited number of cores, with each core possessing a dedicated private cache. The flexibility extends to SIMD instruction sets, with support for architectures like AVX \cite{Lomont2011-qb}, NEON \cite{Reddy2008-gf}, and RVV \cite{Waterman2014-qi}, facilitated through individual vector registers and engines for each core.
\textbf{(a) The LLC} is organized into cache slices interconnected by a Network-on-Chip (NoC), enhancing communication within the system.
Though the depicted model shows four cache slices, it can be expanded to accommodate additional slices as seen in current commercial processors such as the AMD Genoa-X series \cite{Kennedy_undated-bf-2}.
Each cache slice is equipped with a C-SRAM, enabling the rapid retrieval of a full cache block in a single cycle, thereby leveraging the LLC's inherent high bandwidth for compute operations. 
The Data Feeding Module (DFM) within the LLC is managed by the core that retrieves input data (e.g., [1,1024] vector) from the data cache, dispatching it through the NoC. It broadcasts bits to connected C-SRAMs each cycle according to the NBW settings of LUT-based GEMV. A detailed diagram of the DFM, including broadcast logic, the Pattern Reuse Table (PRT), and the adder tree, will be included in the final camera-ready version.
{\textbf{(b) A C-SRAM} is composed of a Bitline Computing SRAM (BC-SRAM), a transpose unit, and a Reconfigurable Control Unit (RCU). The transpose unit's function is to reorient data between horizontal vs.\ vertical layouts, aligning with bit-serial in-SRAM computing requirements \cite{Eckert2018-cl}. The RCU dictates the transpose operation, adjusting it based on the data quantization level, and directs the BC-SRAM's computational processes in response to activation bits. This arrangement allows for efficient data processing directly within the cache hierarchy, reducing data movement overhead and potential bottlenecks.
\textbf{(c) The BC-SRAM} is designed to support the simultaneous activation of two wordlines, which is essential for enabling bitline computing. This is achieved by the inclusion of two decoders.
\textbf{(d) The Sense Amplifiers (SAs)}
are adapted from the Neural Cache structure~\cite{Eckert2018-cl}, employing single-ended SAs to sense wire-AND results from adjacent cells. A lightweight logic stage completes an $n$-bit addition in $n+1$ cycles and an $n$-bit multiplication in $n^2 + 5n - 2$ cycles, where $n$ is the bitwidth. For clarity, the figure omits write-driver circuitry.
\textbf{(e) The transpose unit} converts a horizontal data layout to a vertical one for bit-serial in-SRAM computation. When no AI kernel is active, the C-SRAM can either act as additional LLC capacity or execute simple primitives like compare-and-count for tasks such as log filtering, providing modest latency gains at no extra area cost.

\subsection{Modification on Address Hasher}
Our proposed method requires that weights are evenly distributed across different cache slices, allowing the C-SRAM array to construct LUTs from the nearest data cache slice. This can be achieved by modifying the address hasher. Taking the hasher from \cite{Grohoski2007-dh} as an example, by retaining the lowest 9 bits of the address while scrambling the other bits, it is possible to interleave across all data cache slices at a granularity of 512 bytes.
Moreover, we need to introduce a simple logic within the cache controller to broadcast the input. Upon receiving the \texttt{lutmm\_1k} instruction, this logic should retrieve input data based on the address specified in the $ri$ field and determine the number of bits to broadcast at a time according to the NBW. This ensures that the input vector is efficiently disseminated to the C-SRAM array, enabling parallel construction of multiple LUTs tailored for the subsequent matrix multiplication task.

\subsection{Instruction Execution on Reference Implementation}

To illustrate how our reference implementation works, consider executing the \texttt{lutmm\_k1} instruction
using the left core in Fig.~\ref{fig:overall}(a). 
\textbf{Step 1:} A weight matrix [1024,1024] is fetched from DRAM based on the address in $rw$, and stored in the two cache slices on the left.
\textbf{Step 2:} Utilizing the address in $ri$, the DFM retrieves the [1,1024] input vector from the data cache.
\textbf{Step 3:} Each C-SRAM fetches the weight rows from its data cache that aligns with the NBW setting, governed by the control module within the DFM. The C-SRAM then proceeds to build the LUT using the retrieved weights.
\textbf{Step 4:} The DFM starts broadcasting input data to the connected C-SRAMs, which in turn, execute accumulation and type conversion operations, including partial sum aggregation between two C-SRAMs via the DFM.

\textbf{Step 5:} The intermediate result awaits transmission via NoC to the CPU, where the Vector Engine performs the dequantization and sends the dequantization result back to C-SRAM.

Steps 3 through 5 are repeated $\lceil$1024$/$NBW$\rceil$ times until the multiplication of [1,1024]$\times$[1024,1024] is completed. 
Larger GEMV operations can be realized by repeating the \texttt{lutmm\_1k} instruction, and multiple \texttt{lutmm\_1k} instructions can be executed simultaneously using multiple cores, leveraging our proposed pipeline design to enhance parallelism and throughput.

\section{Evaluation}\label{sec:evaluation}

\subsection{Experimental Setup}
\label{sec:experimental-setup}



\textbf{Architectures and Configurations.}
In our experiments, we consider the following architectures: \emph{ARM CPU}, \emph{AMX CPU}, \emph{GPU}, \emph{Neural Cache}, and \emph{SAIL}. 
~\textbf{ARM CPU.}
This is the baseline CPU only case. We adopt the ARM Neoverse-N1 core with the ARM CMN-600 NoC, based on the Google Cloud Platform (GCP) configuration. The CMN-600 NoC can be scaled to an 8$\times$8 mesh, partitioning a 32\,MB L3 cache into 32 slices of 1\,MB each.
Table~\ref{tab:parameter} summarizes the key parameters of the ARM CPU case.
~\textbf{AMX CPU.}
We use a \texttt{c4-highmem-96} node on GCP, which features a 48-core Intel Emerald Rapids processor with Advanced Matrix Extensions (AMX).
AMX is designed to speed up matrix operations by offloading them to specialized hardware blocks, thus representing the state-of-the-art CPU-based acceleration case.
~\textbf{GPU.}
We conduct GPU-based experiments on a 16-core Intel Skylake CPU machine equipped with four NVIDIA V100 GPUs, provided by GCP. In addition, a single A100 GPU with 80GB HBM2e memory with Intel Gold 6338 GPU was used. 
~\textbf{Neural Cache.}
The Neural Cache architecture is based on the same design as SAIL, with key modifications: LUT-GEMV is replaced by the bit-serial computing method described in \cite{Eckert2018-cl}, and the in-memory type conversion algorithm is excluded.
~\textbf{SAIL.}
SAIL extends \textit{ARM CPU} with 32 Near-Data Processors (NDPs) integrated at the L3 cache, employing existing memory addressing schemes~\cite{Chen2023-jq} to manage cache misses and maintain coherence. The size of each C-SRAM array is 256$\times$512 bits.

\textbf{Benchmarks:} In line with typical cloud-provided LLM inference scenarios, we assume that each configuration serves a single LLM model to multiple users with batching enabled. We benchmark the inference of widely-used, open-source TinyMistral-248M \cite{UnknownUnknown-ab} and Llama-2-7/13B \cite{Touvron2023-qn} model. We have found that llama.cpp (version \texttt{b47879b}) \cite{Gerganov_undated-gt} has the best CPU inference performance, which we use to benchmark both CPU and SAIL results. Specifically, we conduct experiments with a single prompt, single input sequence case, and a \emph{parallel} implementation for batch sizes of 2/4/8, across models quantized at different levels (2/3/4/5/6/8 bits) using the quantization of the llama.cpp framework. We also carry out tests on various platforms to assess performance against a context length of 4096 with different sequence lengths (0 to 2048). We have extended the llama.cpp implementation to support 8-bit quantized KV-cache.

\textbf{Simulator:} We use the gem5 simulator \cite{Binkert2011-ny} for execution-driven, cycle-level simulation, with enhancements to model an NoC and integration of NDPs.
We have calibrated the base out-of-order ARM CPU model of gem5 to match the performance of ARM Neoverse-N1 cores \cite{Arm_Ltd_undated-wi} used in GCP, by (i) utilizing known microarchitectural parameters as listed in 
Table \ref{tab:parameter} and the references cited, and (ii) adjusting unknown parameters such as L2 and L3 cache latencies by comparing with the actual inference latency measurements on GCP servers.
We note that the CPU simulation results using our calibrated model show very good agreement with the actual inference latency on GCP servers, with max difference of 5.4\%.

For NDP simulation, we adopt the methodology used in previous works (e.g., \cite{Zhang2023-rt}), characterizing the cycle counts for key operations such as LUT-based GEMV, batched inference, and in-memory type conversion. Batched inference, in this context, refers to queries that have been pre-organized by inference servers, such as Triton \cite{noauthor_undated-ek} or RayLLM \cite{noauthor_undated-og}, before being fed into the inference engine as batched inputs. Each batched query consists of multiple vectors as input, which are processed together during computation. These cycle numbers, which can be accurately determined with parameters due to the static nature of the computation, are then hardcoded into the NDP model, which is integrated into the gem5 simulator.
%
We have implemented the C-SRAM array using PyMTL3 \cite{Jiang2020-cf} and OpenRAM \cite{Guthaus2016-sw} with FreePDK 45~nm technology \cite{Stine2007-zo}, using Synopsys Design Compiler and Cadence Innovus. The estimated read and write latencies suggest that the C-SRAM array can operate at the system clock frequency of 3~GHz. The area and power estimates are listed in Table \ref{tab:parameter}. 






\begin{table}[tp]
    \centering
    \small
    \caption{System and Architectural Parameters (cyc.: cycle)}
      \label{tab:parameter}
    \scalebox{0.9}{
    \begin{threeparttable}
      \begin{tabular}{cl}
        \toprule
        OOO Cores \cite{Arm_Ltd_undated-wi} & 32 cores, 3 GHz \\
        \midrule
        \multirow{ 3}{*}{Func. Units\cite{neoverse}} & 3 Int ALU \\
        & 2 FP ALU/SIMD \\ 
        & 4-way decode, 8-way issue  \\
        \midrule
        L1 I-Cache & 8 MSHRs, 4-way, 64KB, 1 cyc.\\
        L1 D-Cache & 20 MSHRs, 4-way, 64KB, 3 cyc.\\
        Priv. L2 Cache & 46 MSHRs, 8-way, 1MB, 34 cyc.\\
        \midrule
        Replacement & LRURP\\
        L1/2/3 Stride Pf. & 16 streams, 16 pf./stream\\
        Shared L3 Cache & 128 MSHRs, 16-way, 32MB, 58 cyc., 32 slices\\
        \midrule
        NoC \cite{cmn600} & 32B 1cyc. 8$\times$8 Mesh, 2 GHz\\
        \midrule
        DRAM & 8 channels 3200 MHz DDR4\\
        \midrule
        C-SRAM Array & 256$\times$512 bits.  0.828~mm$^2$, 37.076~mW (est.)\\
        \bottomrule
      \end{tabular}
    \end{threeparttable}
    }
  \vspace{-2em}
  \end{table}
\subsection{Overall Performance}

\begin{table*}[ht]
\centering
\caption{Comparison of Inference Performance Across Quantization Levels and Parallelism on CPU Implementations}
\label{tab:quantization-throughput}
\scalebox{0.8}{
\begin{tabular}{|c|ccc|ccc|ccc|ccc|ccc|ccc|ccc|}
\hline
\textbf{\textsc{Tokens}} & \multicolumn{3}{c|}{\textbf{\textsc{1 Thread}}} & \multicolumn{3}{c|}{\textbf{\textsc{2 Threads}}} & \multicolumn{3}{c|}{\textbf{\textsc{4 Threads}}} & \multicolumn{3}{c|}{\textbf{\textsc{8 Threads}}} & \multicolumn{3}{c|}{\textbf{\textsc{16 Threads}}} \\
\textbf{\textsc{/Second}} & \textbf{\textsc{ARM}} & \textbf{\textsc{AMX}} & \cellcolor[gray]{0.9}\textbf{\textsc{SAIL}} & \textbf{\textsc{ARM}} & \textbf{\textsc{AMX}} & \cellcolor[gray]{0.9}\textbf{\textsc{SAIL}} & \textbf{\textsc{ARM}} & \textbf{\textsc{AMX}} & \cellcolor[gray]{0.9}\textbf{\textsc{SAIL}} & \textbf{\textsc{ARM}} & \textbf{\textsc{AMX}} & \cellcolor[gray]{0.9}\textbf{\textsc{SAIL}} & \textbf{\textsc{ARM}} & \textbf{\textsc{AMX}} & \cellcolor[gray]{0.9}\textbf{\textsc{SAIL}} \\
\hline
\textbf{\textsc{7b-q2}} & 0.68 & 2.06 & \cellcolor[gray]{0.9}\textbf{6.42} & 1.34 & 4.02 & \cellcolor[gray]{0.9}\textbf{12.62} & 2.63 & 7.65 & \cellcolor[gray]{0.9}\textbf{24.00} & 4.97 & 14.25 & \cellcolor[gray]{0.9}\textbf{43.50} & 9.30 & 24.96 & \cellcolor[gray]{0.9}\textbf{81.63} \\
\textbf{\textsc{7b-q3}} & 0.70 & 2.02 & \cellcolor[gray]{0.9}\textbf{5.53} & 1.38 & 3.93 & \cellcolor[gray]{0.9}\textbf{10.93} & 2.71 & 7.47 & \cellcolor[gray]{0.9}\textbf{20.87} & 5.11 & 13.69 & \cellcolor[gray]{0.9}\textbf{38.40} & 9.62 & 24.50 & \cellcolor[gray]{0.9}\textbf{73.75} \\
\textbf{\textsc{7b-q4}} & 0.70 & 3.45 & \cellcolor[gray]{0.9}\textbf{4.82} & 1.37 & 6.72 & \cellcolor[gray]{0.9}\textbf{9.61} & 2.67 & 11.51 & \cellcolor[gray]{0.9}\textbf{18.67} & 5.15 & 21.13 & \cellcolor[gray]{0.9}\textbf{35.17} & 9.85 & 33.55 & \cellcolor[gray]{0.9}\textbf{72.10} \\
\textbf{\textsc{(Intel-q4)}} &  & (2.32) & \cellcolor[gray]{0.9} &  & (4.39) & \cellcolor[gray]{0.9} &  & (8.22) & \cellcolor[gray]{0.9} &  & (14.42) & \cellcolor[gray]{0.9} &  & (22.49) & \cellcolor[gray]{0.9} \\
\textbf{\textsc{7b-q5}} & 0.60 & 1.30 & \cellcolor[gray]{0.9}\textbf{3.98} & 1.17 & 2.56 & \cellcolor[gray]{0.9}\textbf{7.96} & 2.32 & 4.84 & \cellcolor[gray]{0.9}\textbf{15.52} & 4.48 & 9.17 & \cellcolor[gray]{0.9}\textbf{29.62} & 8.49 & 16.48 & \cellcolor[gray]{0.9}\textbf{61.84} \\
\textbf{\textsc{7b-q6}} & 0.79 & 1.20 & \cellcolor[gray]{0.9}\textbf{3.34} & 1.20 & 2.33 & \cellcolor[gray]{0.9}\textbf{6.67} & 2.36 & 4.47 & \cellcolor[gray]{0.9}\textbf{12.97} & 4.52 & 8.10 & \cellcolor[gray]{0.9}\textbf{24.60} & 8.31 & 14.62 & \cellcolor[gray]{0.9}\textbf{50.63} \\
\textbf{\textsc{7b-q8}} & 0.66 & 2.30 & \cellcolor[gray]{0.9}\textbf{2.60} & 1.28 & 4.51 & \cellcolor[gray]{0.9}\textbf{5.22} & 2.51 & 7.50 & \cellcolor[gray]{0.9}\textbf{10.28} & 4.69 & 13.55 & \cellcolor[gray]{0.9}\textbf{19.86} & 5.54 & 18.39 & \cellcolor[gray]{0.9}\textbf{43.27} \\
\textbf{\textsc{(Intel-q8)}} &  & (1.60) & \cellcolor[gray]{0.9} &  & (3.09) & \cellcolor[gray]{0.9} &  & (5.80) & \cellcolor[gray]{0.9} &  & (10.32) & \cellcolor[gray]{0.9} &  & (15.59) & \cellcolor[gray]{0.9} \\
\hline
\textbf{\textsc{13b-q2}} & 0.35 & 1.06 & \cellcolor[gray]{0.9}\textbf{3.77} & 0.70 & 2.06 & \cellcolor[gray]{0.9}\textbf{7.44} & 1.38 & 3.91 & \cellcolor[gray]{0.9}\textbf{14.34} & 2.68 & 7.28 & \cellcolor[gray]{0.9}\textbf{26.63} & 5.05 & 12.75 & \cellcolor[gray]{0.9}\textbf{52.55} \\
\textbf{\textsc{13b-q3}} & 0.35 & 1.02 & \cellcolor[gray]{0.9}\textbf{3.67} & 0.69 & 2.01 & \cellcolor[gray]{0.9}\textbf{7.33} & 1.36 & 3.82 & \cellcolor[gray]{0.9}\textbf{13.84} & 2.63 & 7.00 & \cellcolor[gray]{0.9}\textbf{25.70} & 5.01 & 12.62 & \cellcolor[gray]{0.9}\textbf{51.10} \\
\textbf{\textsc{13b-q4}} & 0.36 & 1.82 & \cellcolor[gray]{0.9}\textbf{2.81} & 0.72 & 3.53 & \cellcolor[gray]{0.9}\textbf{5.62} & 1.41 & 5.79 & \cellcolor[gray]{0.9}\textbf{11.00} & 2.75 & 10.95 & \cellcolor[gray]{0.9}\textbf{21.06} & 5.27 & 17.42 & \cellcolor[gray]{0.9}\textbf{45.07} \\
\textbf{\textsc{(Intel-q4)}} &  & (1.21) & \cellcolor[gray]{0.9} &  & (2.31) & \cellcolor[gray]{0.9} &  & (4.39) & \cellcolor[gray]{0.9} &  & (7.87) & \cellcolor[gray]{0.9} &  & (12.74) & \cellcolor[gray]{0.9} \\
\textbf{\textsc{13b-q5}} & 0.31 & 0.67 & \cellcolor[gray]{0.9}\textbf{2.32} & 0.61 & 1.32 & \cellcolor[gray]{0.9}\textbf{4.64} & 1.20 & 2.52 & \cellcolor[gray]{0.9}\textbf{9.10} & 2.34 & 4.78 & \cellcolor[gray]{0.9}\textbf{17.60} & 4.44 & 8.56 & \cellcolor[gray]{0.9}\textbf{38.24} \\
\textbf{\textsc{13b-q6}} & 0.32 & 0.62 & \cellcolor[gray]{0.9}\textbf{1.94} & 0.62 & 1.18 & \cellcolor[gray]{0.9}\textbf{3.88} & 1.23 & 2.17 & \cellcolor[gray]{0.9}\textbf{7.60} & 2.40 & 4.14 & \cellcolor[gray]{0.9}\textbf{14.61} & 4.52 & 7.25 & \cellcolor[gray]{0.9}\textbf{31.32} \\
\textbf{\textsc{13b-q8}} & 0.34 & 1.15 & \cellcolor[gray]{0.9}\textbf{1.51} & 0.68 & 2.20 & \cellcolor[gray]{0.9}\textbf{3.03} & 1.29 & 3.89 & \cellcolor[gray]{0.9}\textbf{5.98} & 2.46 & 7.19 & \cellcolor[gray]{0.9}\textbf{10.75} & 4.80 & 10.07 & \cellcolor[gray]{0.9}\textbf{26.25} \\
\textbf{\textsc{(Intel-q8)}} &  & (0.83) & \cellcolor[gray]{0.9} &  & (1.62) & \cellcolor[gray]{0.9} &  & (3.06) & \cellcolor[gray]{0.9} &  & (5.57) & \cellcolor[gray]{0.9} &  & (8.56) & \cellcolor[gray]{0.9} \\
\hline
\textbf{\textsc{Geo-Mean}} & \textbf{0.52} & \textbf{1.31} & \cellcolor[gray]{0.9}\textbf{3.50} & \textbf{1.02} & \textbf{2.58} & \cellcolor[gray]{0.9}\textbf{6.97} & \textbf{2.00} & \textbf{4.74} & \cellcolor[gray]{0.9}\textbf{13.49} & \textbf{3.85} & \textbf{8.67} & \cellcolor[gray]{0.9}\textbf{25.37} & \textbf{7.20} & \textbf{14.67} & \cellcolor[gray]{0.9}\textbf{52.24} \\
\hline
\end{tabular}}
\footnotesize
\begin{flushleft}
* ARM and SAIL results are from gem5 simulation while AMX is from real hardware execution.  ARM results are validated with real hardware results. \\
** AMX results are based on llama.cpp \cite{Gerganov_undated-gt}, except those in parentheses, which are based on Intel-extension-for-PyTorch \cite{UnknownUnknown-bx} supporting Q4 and Q8 only.
\end{flushleft}
\vspace{-2em}
\end{table*}

Table \ref{tab:quantization-throughput} presents the throughput for 7B and 13B models across various quantization levels and thread counts on three architectures. The results are based on \textit{llama.cpp}~\cite{Gerganov_undated-gt}, except for the numbers in parentheses, which use the \textit{Intel-extension-for-PyTorch}~\cite{UnknownUnknown-bx}.

SAIL's performance scales significantly better with thread count compared to the other architectures. For instance, in the 7B-Q8 case, ARM's 16-thread per-thread performance drops to 54\% of its single-thread performance, while SAIL maintains 87\%. AMX's scalability is intermediate at 61\%. This improved scalability is primarily due to SAIL's architecture mitigating cache contention and synchronization overhead that affects traditional multi-core CPUs during memory-intensive LLM inference. Consequently, while ARM and AMX show sublinear scaling, SAIL achieves near-linear performance growth.

\subsection{Sensitivity to Quantization Level}

\begin{figure}[tp]
\centerline{\includegraphics[width=3.0in]{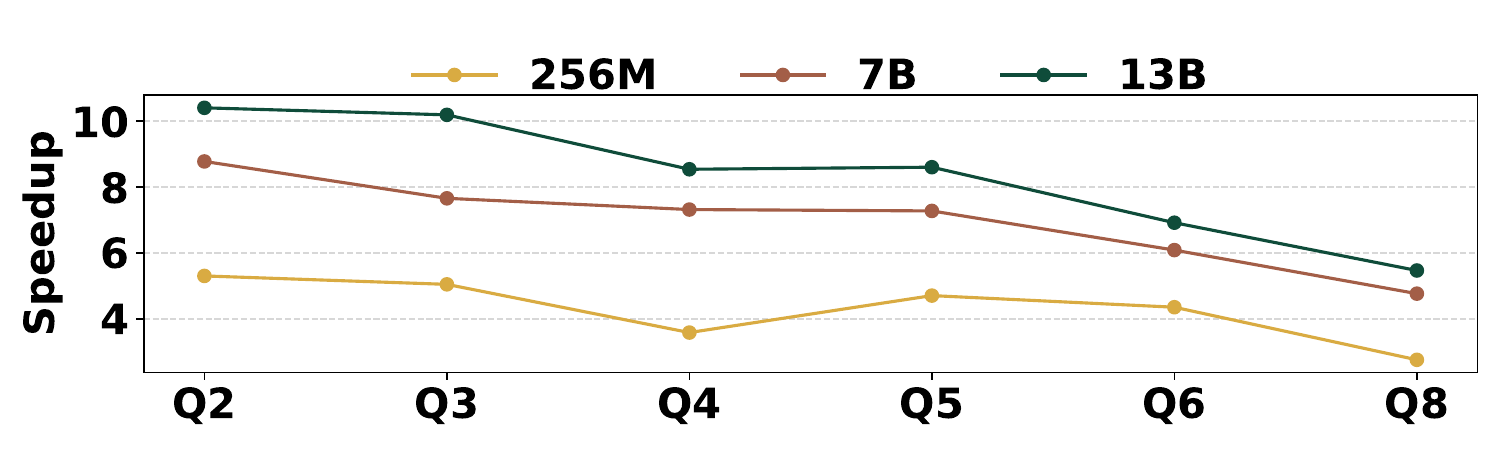}}
\caption{
Speedups of SAIL over models with different quantization levels.
}
\label{fig:eval-3}
\vspace{-1em}
\end{figure}

Fig.~\ref{fig:eval-3} shows SAIL’s token generation speed across various quantization levels. SAIL consistently outperforms ARM, with the advantage being most pronounced at lower precisions due to the efficiency of LUT-based GEMV. For example, at Q2, SAIL achieves speedups of up to 10.41$\times$ on the 13B model. While the performance gap narrows at higher bit-widths as conventional CPUs benefit from native int4/int8 support, SAIL still maintains a significant advantage. Larger, more memory-bound models like the 13B benefit more from SAIL's near-cache design, which minimizes data movement.

\subsection{Sensitivity to Batch Size}

\begin{figure}[tp]
\centerline{\includegraphics[width=3.0in]{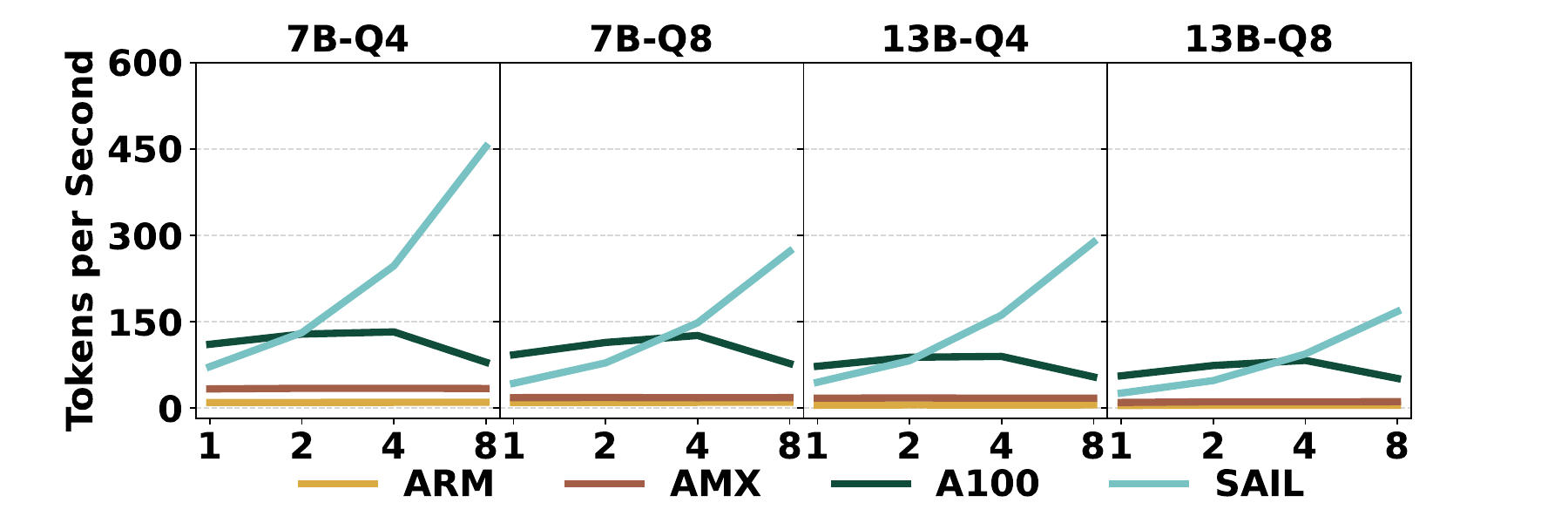}}
\caption{
Token generation speeds of different platforms for various models and batch sizes.
}
\label{fig:eval-7}
\vspace{-1em}
\end{figure}


Fig.~\ref{fig:eval-7} shows the token generation speed for batched inference. We used 16 threads for all CPU-based architectures and benchmarked against a single A100 GPU using the highly optimized \textit{llama.cpp} framework. Continuous batching optimizations, like those in vLLM, are orthogonal and could benefit all architectures, including SAIL.

While batching is common for GPUs, limited VRAM restricts the maximum batch size, as seen with the A100 on the 13B-Q8 model. CPU-based platforms like ARM and AMX show minimal benefit from batching due to memory bandwidth saturation. In contrast, SAIL benefits the most, leveraging its LUT-GEMV technique, tensor-level scheduling, and pipeline design. For the 7B-Q4 model, SAIL achieves a 13.2$\times$ speedup over AMX and a 3.42$\times$ speedup over the A100 GPU. The performance gap narrows for larger models, but SAIL still maintains a 2.02$\times$ speedup on the 13B-Q8 model.

\subsection{Comparison with Latest CPU Baselines}

\begin{figure}[tp]
 \centerline{\includegraphics[width=3.0in]{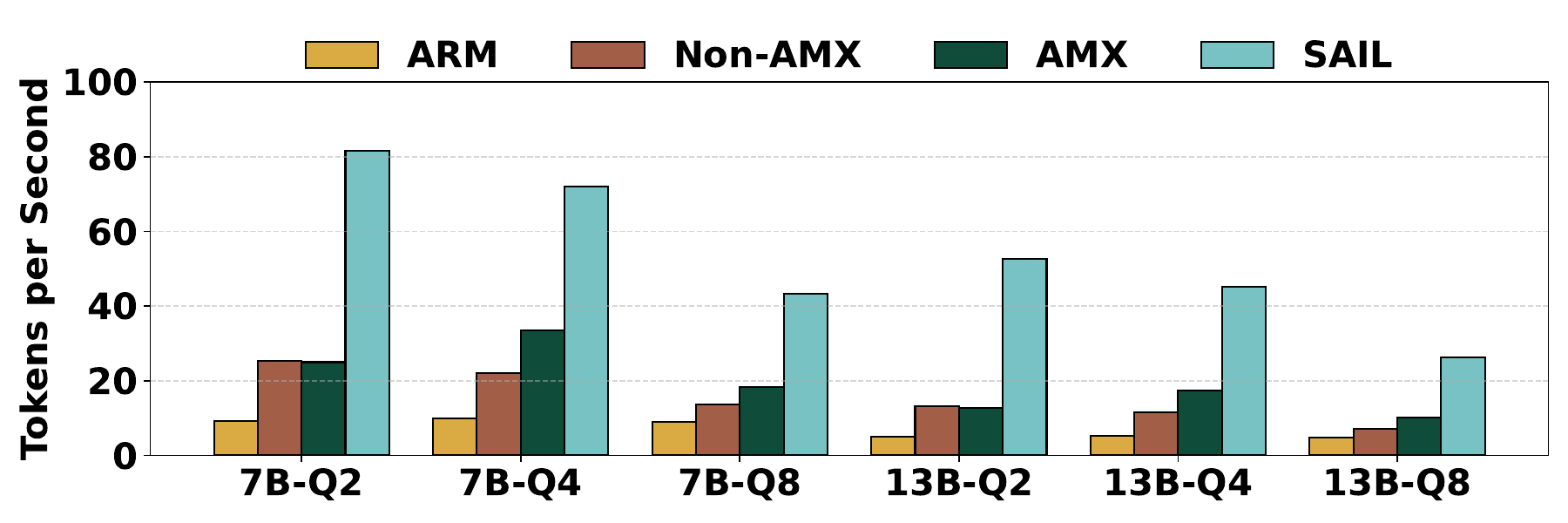}}
\caption{Performance comparison of ARM, Non-AMX, AMX, and SAIL at 7B and 13B under Q2, Q4, and Q8.
}
\vspace{-1em}
\label{fig:eval-2}
\end{figure}
 
Figure~\ref{fig:eval-2} compares the throughput (in tokens/s) of ARM, Non-AMX, AMX, and SAIL on 7B and 13B with Q2, Q4, and Q8. SAIL consistently outperforms all CPU-based approaches, such as achieving 81.63~tokens/s at 7B-Q2 versus around 25~tokens/s for AMX and Non-AMX. 
The AMX hardware only supports int8 and BF16 data types. As a result, the benefits of AMX's specialized instructions diminish at Q2, where Non-AMX and AMX offer similar performance. At Q4 and Q8, AMX begins to outperform Non-AMX due to native INT8 acceleration, yet still trails SAIL. Similarly, the NPUs in recent CPUs like Lunar Lake and Strix Halo are limited to INT8 or FP16, making them unsuitable for LLMs that rely on sub-8-bit precision. The LUT-GEMV method of SAIL natively supports arbitrary bit-widths, making it faster at all quantization levels.

\subsection{Performance Breakdown}

\begin{figure}[tp]
\centerline{\includegraphics[width=3.0in]{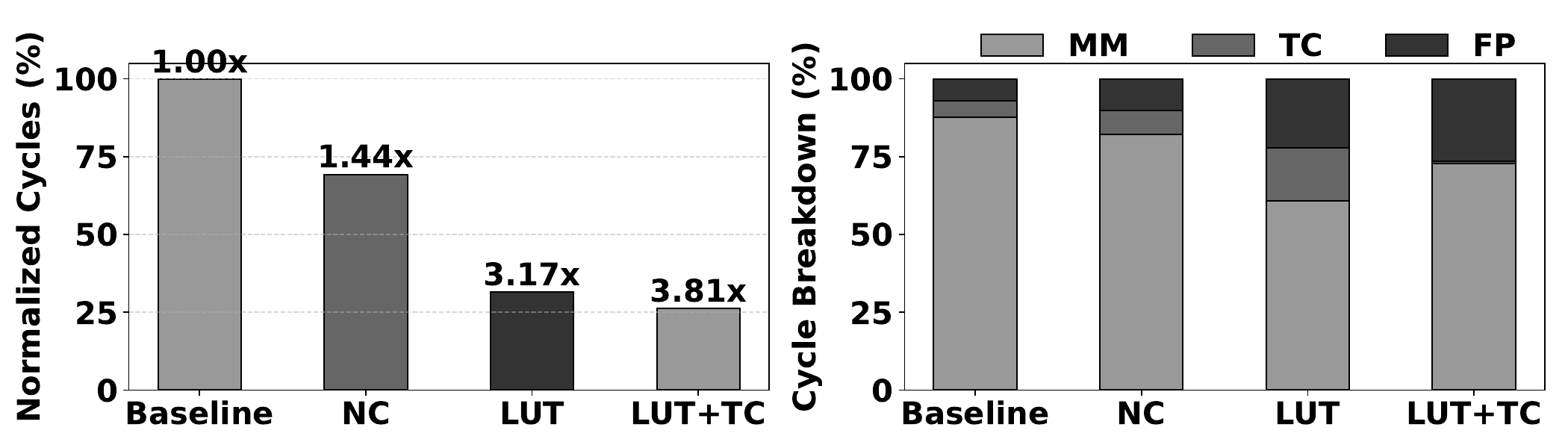}}
\caption{
Performance breakdown of the proposed in-memory LUT-based GEMV and in-memory type conversion approach, compared with the baseline CPU and Neural Cache (NC) implementations.
}
\label{fig:ablation}
\vspace{-1em}
\end{figure}

Figure~\ref{fig:ablation} shows a latency breakdown for a Q4 GEMV kernel, comparing four configurations: a real ARM machine (\emph{Baseline}), \emph{NC} (Neural Cache), \emph{LUT} (SAIL without in-memory type casting), and \emph{LUT+TC} (full SAIL). NC provides a speedup over the Baseline by moving computation closer to data. LUT-GEMV further improves performance by converting multiplications into cheaper additions via a reusable lookup table. Finally, our in-memory type conversion algorithm parallelizes data-format changes, reducing latency even more and achieving a final 3.81$\times$ speedup over the Baseline.



\subsection{Comparison with GPU}  
Here we compare the absolute performance of SAIL and GPU inference. 
We compare SAIL with 16 threads and the batch size of 8 vs.\ NVIDIA V100 GPUs. Since the VRAM capacity of GPUs significantly impacts the performance, we also compare SAIL with 2$\times$V100, 32~GB VRAM in total and a single A100 GPU with 80GB of HBM2e memory. Note that increasing the number of GPUs does not noticeably increase the performance, but it does enable a larger model and/or larger context length. To make a fair comparison with SAIL results, the same llama.cpp framework was used to benchmark GPU performance. Although there exists HuggingFace \cite{wolf-etal-2020-transformers}, a more popular LLM inference framework for GPUs, we found that the latest version of llama.cpp (version \texttt{e09a800}, which is used for GPU experiments only) provides better throughput for quantized inference compared to HuggingFace. 
The continuous batching \cite{Kwon2023-bh} is considered an orthogonal method, since it can benefit all KV-cache-enabled solutions including SAIL.

Table \ref{tab:gpu-comparison} compares the inference throughput to generate 512 tokens for various model sizes and quantization levels, only for the generation stage of LLM inference. The cases where SAIL performs better than the GPU(s) are highlighted. For the SAIL case, we observe that the inference throughout is not affected by the context length, so only the results with the context length of 4096 is reported (the default Llama-2 context length). In the case of GPU, the choice of context length determines the maximum batch size, which then impacts the throughput. Therefore, we have tested various batch sizes and report the best performing case. In the case of V100 results, increasing the batch size over 8 did not increase the inference throughput although a higher batch size of 16 or 32 was tested. 
We observe that for both model sizes, SAIL performs better than V100 GPUs for context lengths 1K and above. Note that the capability to use longer context length is crucial in LLM inference, as it directly impacts the model performance to capture more information from the input prompt. 
Therefore, the SAIL's ability to efficiently support longer context lengths is an important advantage of our proposed approach. SRAM-PIM inside GPUs offers limited benefit because their bottleneck is off-chip HBM bandwidth; CPU caches remain a better integration point for PIM.

\definecolor{lightblue}{RGB}{230, 240, 250}
\definecolor{mediumblue}{RGB}{200, 220, 240}
\definecolor{darkblue}{RGB}{170, 200, 230}

\definecolor{lightgray}{RGB}{240, 240, 240}
\definecolor{mediumgray}{RGB}{220, 220, 220}
\definecolor{darkgray}{RGB}{200, 200, 200}

\definecolor{lightorange}{RGB}{255, 248, 230}
\definecolor{mediumorange}{RGB}{255, 240, 200}
\definecolor{darkorange}{RGB}{255, 230, 170}

\begin{table}[t!]
\centering
\small
\scalebox{0.9}{
\begin{threeparttable}
\caption{Token Generation Speed Comparison with GPUs}
\label{tab:gpu-comparison}
\begin{tabular}{cc|cc|cc}
\toprule
\multicolumn{2}{c|}{}            & \multicolumn{4}{c}{Tokens per second / batch size} \\
\midrule
\multicolumn{2}{c|}{Model}            & \multicolumn{2}{c|}{Llama-2-7B}                                                                                         & \multicolumn{2}{c}{Llama-2-13B}                                                                                        \\
\multicolumn{2}{c|}{Quantization Level} & \multicolumn{1}{c}{Q4} & \multicolumn{1}{c|}{Q8} & \multicolumn{1}{c}{Q4} & \multicolumn{1}{c}{Q8} \\
\midrule
\multicolumn{2}{c|}{Context Length} & \multicolumn{1}{c}{} & \multicolumn{1}{c|}{} & \multicolumn{1}{c}{} & \multicolumn{1}{c}{} \ \\
\multirow{4}{*}{\makecell{1xV100}} 
& 512 &216.3 / 8& 190.5 / 8&  173.9 / 8& 80.02 / 4\\
& 1K & \cellcolor{lightblue}173.4 / 4&   \cellcolor{lightblue}126.9 / 4& 126.4 / 4&  \cellcolor{lightblue}42.01 / 2\\
& 2K & \cellcolor{mediumblue}123.6 / 2&   \cellcolor{mediumblue}84.98 / 2& \cellcolor{mediumblue}85.47 / 2 & \cellcolor{mediumblue}38.27 / 1\\
& 4K &  \cellcolor{darkblue}78.98 / 1&   \cellcolor{darkblue}41.62 / 1&\cellcolor{darkblue}39.97 / 1 &  \multicolumn{1}{c}{\cellcolor{gray!30}\ding{55}}\\
\midrule
\multirow{4}{*}{\makecell{2xV100}} 
& 512& 229.3 / 8& 196.3 / 8 & 148.5 / 8& 127.1 / 8\\
& 1K & \cellcolor{lightblue}179.6 / 8&  163.3 / 8& 114.7 / 8& 104.5 / 8\\
& 2K & \cellcolor{lightblue}129.7 / 4& \cellcolor{lightblue}112.6 / 8&\cellcolor{lightblue}81.99 / 4 & \cellcolor{lightblue}67.71 / 4 \\
& 4K &  \cellcolor{mediumblue}88.02 / 4&  \cellcolor{mediumblue}81.90 / 8 & \cellcolor{mediumblue}51.15 / 2& \cellcolor{mediumblue}44.68 / 2\\
\midrule
\multirow{4}{*}{\makecell{A100}} 
& 512& 670.7 / 32& 652.4 / 32 & 442.4 / 32& 431.0 / 32\\
& 1K & 425.8 / 32&  418.2 / 32& 278.8 / 32& 274.4 / 32\\
& 2K & 255.8 / 32& 252.7 / 32& 117.9 / 8 & 109.7 / 8 \\
& 4K &  \cellcolor{lightblue}129.3 / 4&  \cellcolor{lightblue}120.4 / 4 & \cellcolor{lightblue}87.50 / 4& 80.16 / 4\\
\midrule
\rowcolor{lightblue}
\multicolumn{2}{c|}{\textbf{SAIL-16T-8B}} & \textbf{199.28 / 8} & \textbf{134.22 / 8} & \textbf{113.84 / 8} & \textbf{73.93 / 8} \\
                     \bottomrule
\multicolumn{6}{l}{\small \ding{55} does not fit in GPU memory. Numbers after slash are best batch sizes.}
\end{tabular}
\end{threeparttable}
}
\vspace{-1em}
\end{table}
\begin{figure*}[htp]
\centerline{\includegraphics[width=6.0in]{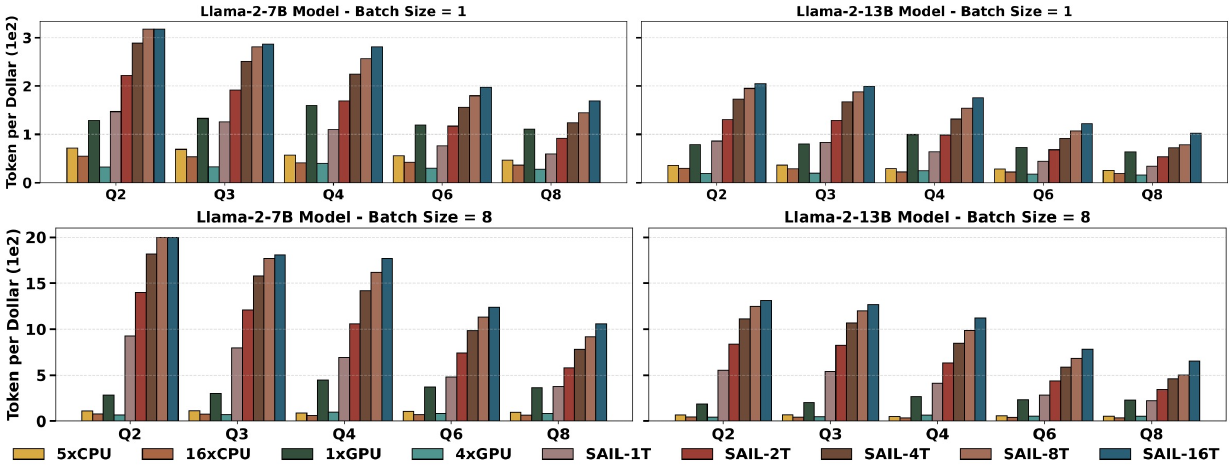}}
\caption{
Tokens per Dollar over different models and quantization levels across different platforms, with batch size 1 and 8. 
}
\label{fig:eval-6}
\vspace{-1em}
\end{figure*}

\subsection{Tokens per Dollar (TPD) across Platforms}
\label{tpd-def}


We measure cost-effectiveness through the \emph{Tokens per Dollar (TPD)} metric, calculated as TPD = (tokens/s $\times$ 30 days) / (monthly on-demand GCP price). This metric combines each platform’s observed throughput with its estimated monthly cost from Table~\ref{tab:price_parameters}, effectively folding capital expenditure (CAPEX), energy, and operational expenditure (OPEX) into a single, user-visible cost. Figure~\ref{fig:eval-6} presents TPD results for the 7B and 13B models at quantization levels Q8/6/4/3/2, spanning CPU-based (5-core and 16-core), single- and multi-GPU (V100), and SAIL configurations.

At moderate quantizations (Q8--Q3), a single V100 GPU generally achieves higher TPD than a single-thread SAIL configuration, reflecting the GPU’s substantial raw compute power in relation to its cost. However, when the bitwidth drops to Q2, SAIL-1T surpasses the GPU in TPD, revealing that the GPU is comparatively less efficient at ultra-low-bit operations. SAIL’s LUT-based near-data approach, by contrast, retains robust throughput under low-precision settings. This advantage persists when scaling up from 7B to 13B, suggesting that even for larger model sizes, SAIL’s specialized low-bit hardware offers favorable cost-effectiveness.

In addition to quantization effects, the data also illustrate the influence of batch size and thread count. Increasing the batch size from 1 to 8 enhances TPD across all platforms, but the degree of improvement varies. Multi-GPU setups see performance gains but suffer from steep hardware and operational costs, curtailing their overall TPD growth. CPU-based solutions also exhibit some improvement but often encounter memory bandwidth bottlenecks. As the batch size increases to 8, SAIL surpasses the single-GPU configuration in TPD across all tested quantization levels except for the 13B Q8 setting under a single-thread mode. This improvement stems from amortizing the LUT construction overhead across a greater number of tokens, thereby reducing the effective memory bandwidth demand and improving parallel utilization.
\begin{table}[tp]
  \centering
  \small
  \caption{Cost Estimation Based on GCP \cite{UnknownUnknown-bl}}
  \label{tab:price_parameters}
  \scalebox{0.9}{
  \begin{threeparttable}
    \begin{tabular}{cc}
      \toprule
      System  & Monthly price (\$)  \\
      \midrule 
 
      5-core CPU  w/ 32 GB DRAM & 292.31 \\ 
      16-core CPU w/ 32 GB DRAM & 665.45 \\ 
      2-core CPU 1xV100 GPU$ \dag  $ w/ 15GB DRAM & 1861.5 \\ 
      2-core CPU 4xV100 GPU$ \dag  $ w/ 15GB DRAM & 7446.0 \\ 
      \bottomrule
      \multicolumn{2}{l}{$ \dag  $16GB VRAM for each GPU}
    \end{tabular}
  \end{threeparttable}}
  \vspace{-1em}
\end{table}

\subsection{Hardware Overhead} \label{sec:hardware-overhead}

In this section, we delve into the hardware overhead required for our proposed architecture, starting with the C-SRAM capacity. As per our design, each thread is capable of managing a [1,1024]$\times$[1024,1024] GEMV computation, which necessitates control over two C-SRAM blocks of size 256$\times$512 (bits), amounting to 32KB of SRAM per thread. In our experiments, we scaled up to 16 threads, resulting in an additional 512KB of C-SRAM. This incremental overhead is marginal---only about 1.6\%---compared with our 32MB LLC. Referencing the previous work \cite{Al-Hawaj2023-rr}, the energy cost for C-SRAM is around 20\%, and the area overhead is about 10\%---at the SRAM level. The overhead at the system level is much lower (see the next section). 

\subsection{Overhead Comparison with ASICs and PIMs 
}




Table~\ref{tab:comparison} summarizes the hardware (HW) and system (Sys.) overhead of large-scale ASICs (e.g., TPU~\cite{Jouppi2017-zp}), small-scale ASICs (e.g., AMX~\cite{Kim2024-ja}), PIMs (e.g., EVE~\cite{Al-Hawaj2023-rr}), and our proposed SAIL. Large-scale ASICs often rely on extensive on-chip buffers, specialized logic, and external DRAM memory which is not scalable, while small-scale ASICs like AMX introduce accelerator blocks for tile-based matrix multiplication and require tile-specific instruction and compiler support. PIMs embed compute peripherals in the memory arrays with an area overhead of around 10\%, necessitating OS-level modifications for managing near-data operations. By contrast, SAIL adds only minimal logic to the CPU cache, introduces a single LUT-based instruction, and works seamlessly with the standard memory hierarchy, resulting in around 2\% area overhead.

\begin{table}[tp]
\centering
\small
\caption{Overhead Comparison of Different Architectures}
\label{tab:comparison}
\scalebox{0.9}{
\begin{threeparttable}
\begin{tabular}{p{1.7cm} p{2.6cm} p{2.6cm}}
\toprule
\textbf{Approach} & \textbf{HW Overhead} & \textbf{Sys. Overhead}\\
\midrule
\textbf{Large-scale ASICs} \cite{Jouppi2017-zp} & 
Large buffers and dedicated logics & 
Limited memory scalability\\
\midrule
\textbf{Small-scale ASICs} \cite{Kim2024-ja} & 
Extra accelerator for tile-based MM & 
Special instructions and compiler \\
\midrule
\textbf{PIMs} \cite{Al-Hawaj2023-rr} & 
Compute peripherals ($\sim$10\% area) & 
New instructions \& OS modifications \\
\midrule
\textbf{SAIL} & 
Minimal CPU and cache modifications ($\sim$2\% area)\tnote{1} & 
Only One instruction; standard memory hierarchy \\
\bottomrule
\end{tabular}
\begin{tablenotes}
\item[1] Includes DFM, adder tree, and pattern redundancy logic; each contributes <0.1\% of a 1MB cache slice area.
\end{tablenotes}
\end{threeparttable}
}
\vspace{-1em}
\end{table}

\section{Conclusion}

We present SAIL (SRAM-Accelerated Inference of Large Language Models), a parallel inference system that boosts batched input efficiency via tensor-level scheduling and in-SRAM LUT-based GEMV with pattern-aware optimization. By combining near-cache-slice and in-SRAM computing, SAIL overcomes prior PIM-based limitations, delivering efficient inference across diverse batch sizes, model scales, and quantization levels. SAIL achieves notable performance and cost benefits over CPU, GPU, and existing PIM platforms.


\bibliographystyle{IEEEtranS}
\bibliography{1-ref,references,paperpile}

\end{document}